\documentstyle[11pt,magfields_pasp,twoside,epsf]{article}
\def\edcomment#1{\iffalse\marginpar{\raggedright\sl#1\/}\else\relax\fi}
\reversemarginpar

\def\kms{\hbox{km$\,$s$^{-1}$}}

\def\Ilambda{\hbox{I$_{\lambda}$}}
\def\Flam{\hbox{F$_{\lambda}$}}
\def\1half{\hbox{\tiny 1/2}}
\def\3half{\hbox{\tiny 3/2}}

\begin{document}
\title{Determination of Magnetic Fields in the Winds from Hot Stars Using
the Hanle Effect}
\author{Joseph P. Cassinelli$^1$, Kenneth H. Nordsieck$^1$, \& Richard
Ignace$^2$}
\affil{$~^1$ Astronomy Dept. Univ. of Wisconsin, 475 N. Charter St. 
Madison WI. 53706, USA}
\affil{$~^2$ Dept. of Physics and Astronomy, Univ. of Iowa, 203 Van
Allen Hall, Iowa City IA, 52242, USA}

\begin{abstract}
Resonance lines that are sensitive to the Hanle effect are prominent in the
UV spectra of early-type stars. To understand the differences from the solar
application of the Hanle effect, we focus on the formation of P-Cygni lines
both as a scattering process, and as one that allows a spectral isolation of
sectors in the wind. Some complications occuring in the solar case are found
to be absent for the Hanle effect for hot stars. Rocket observations 
from the Far Ultraviolet SpectroPolarimeter (FUSP) experiment should allow 
for a determination of fields in the dynamically interesting range from 1 
to 300 Gauss.
\end{abstract}

\section{Introduction}

The Hanle effect has never been measured for any star other than the Sun,
but it should be observable in the UV spectra of hot-stars such as O stars,
OB supergiants, Wolf Rayet Stars and Be stars. These stars have
strong stellar winds driven, in part, by the same resonance
lines that are sensitive to the Hanle effect. The observed lines are
have full widths of about 4000 to 5000 \kms, and are commonly resolved by
satellite UV observatories. In the case of the rapidly
rotating emission line Be stars, and in some Wolf Rayet stars, observations
of the polarization owing to electron scattering have been made from space,
but at a lower resolution than is needed for Hanle studies. 

\section{Hot Star Line Profile Formation}

For the Sun, observers can measure the specific intensity, \Ilambda, of the
light. Other stars are point-like so only the flux, \Flam, can be
measured, and this is an average of the intensity across the face of the
star and its envelope. Ignace et al. (1997, 1999, 2001) have addressed this
problem of what can be learned from polarized fluxes, and have shown that
useful information can be derived about both the field strength and 
the magnetic geometry in winds from hot stars.

The Hanle diagnostic is possible because the lines are formed by a
scattering process, and the lines profiles also allow us to isolate spatial
sectors of the envelope of the star. To illustrate the scattering nature of
the line formation process we use the ``spherical shell'' explanation of
P-Cygni lines. To describe the spatial sectoring, we use the ``iso-velocity'' 
or constant line of sight velocity approach.

\begin{figure}[h]
\plottwo{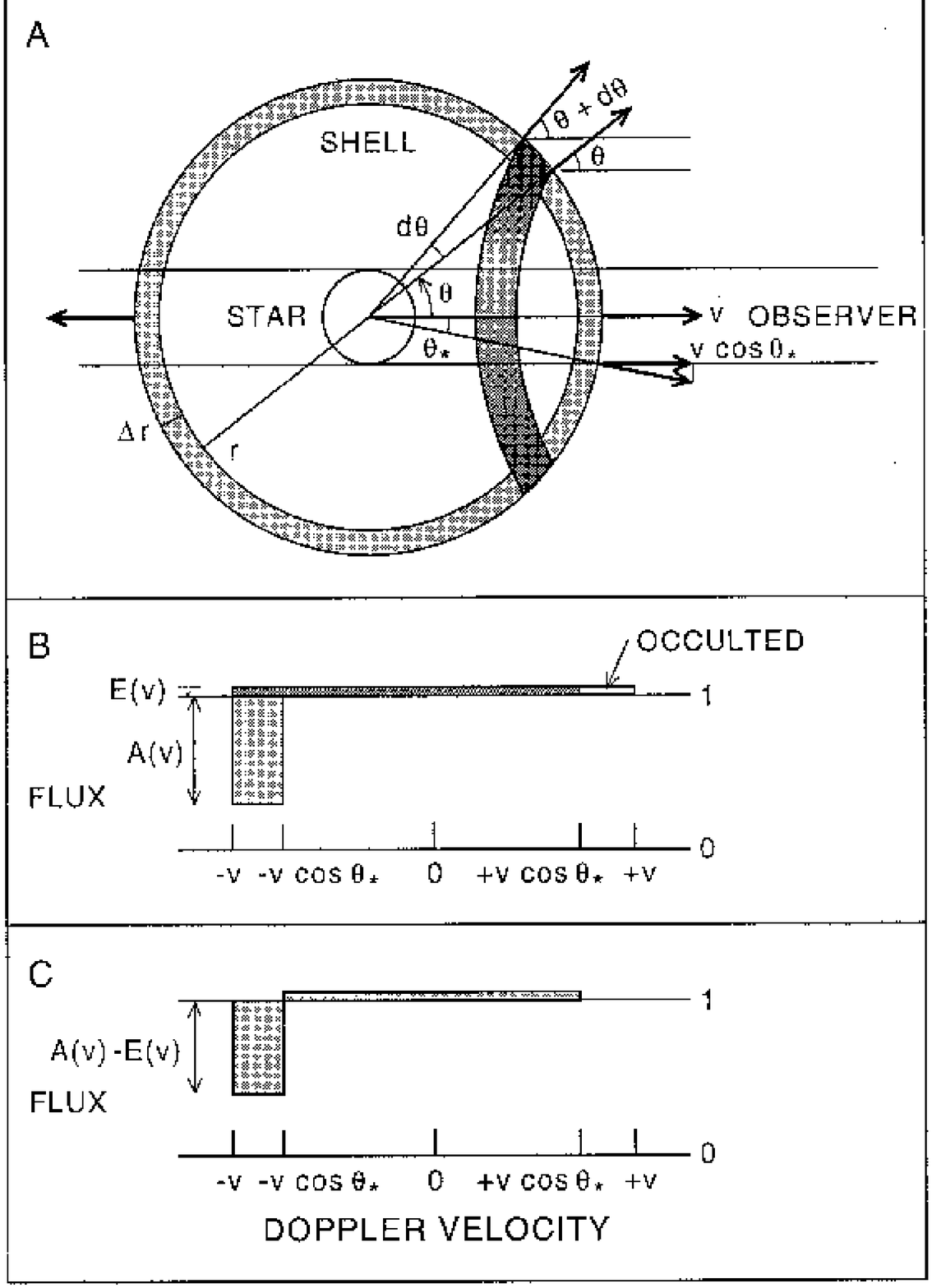}{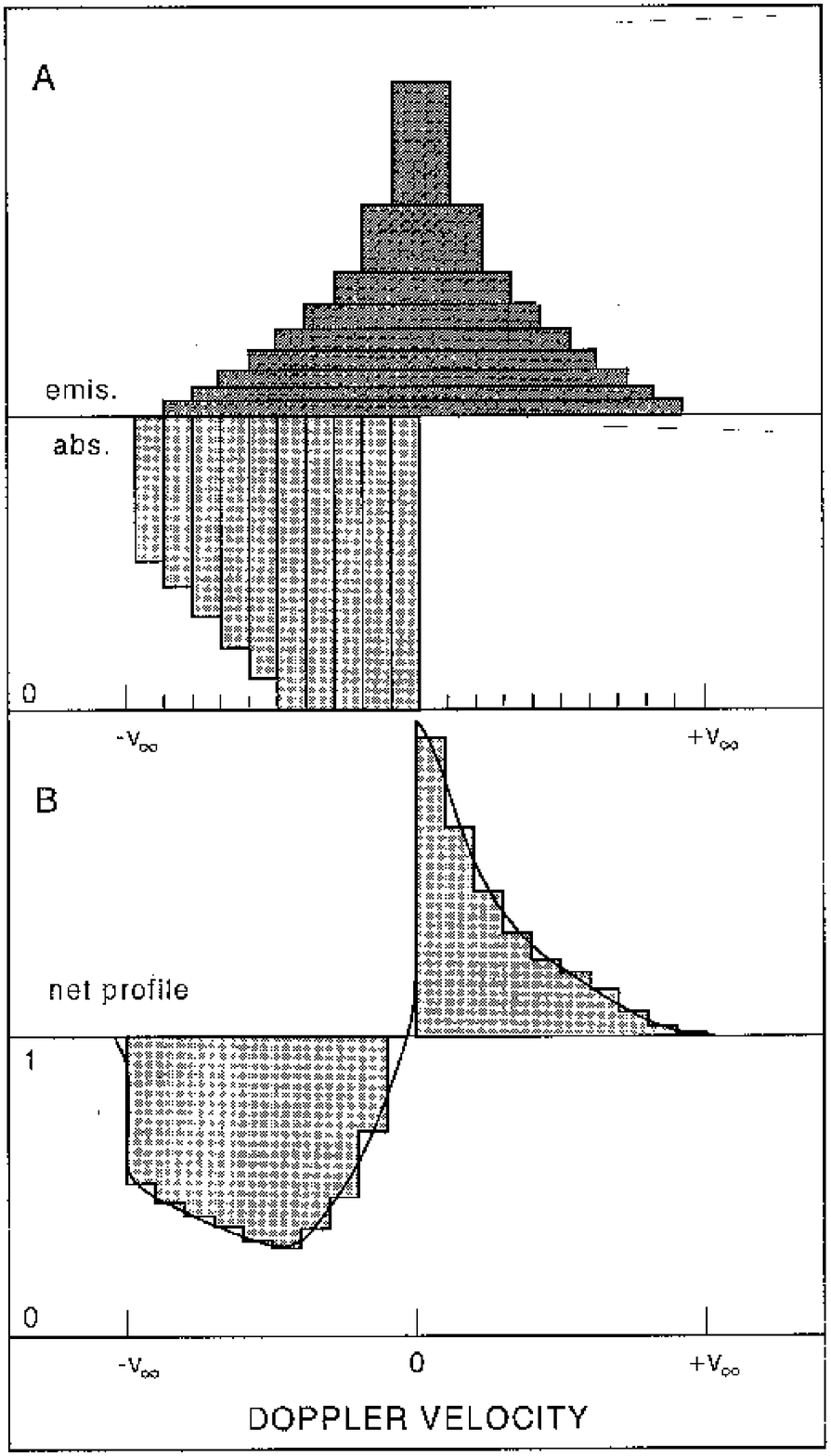}
\caption{The shell model for the formation of a P-Cygni lines by a purely 
scattering process.  Each spherical shell gives rise to a thin
``absorption'' component where light is scattered away from the observer,
The scattering of the same amount of light toward the observer 
by the rest of the shell and produces a flat emission component. 
(adapted from Lamers and Cassinelli, 1999)}
\end{figure}
The left panel of Figure 1, we see that the line is composed of narrow boxes
of attenuated light plus broad flat topped emission components. The entire
profile is composed of the sum of many components as illustrated in the
right panel. The P-Cygni line profile is formed entirely by a {\it
scattering} process. The emergent scattered light is distributed across the
line, even in the so-called absorption side of the line. Using this
spherical shell approach to interpret line profiles, modelers regularly
derive velocity and ionization structures of hot star winds.

\begin{figure}[ht]
\plotone{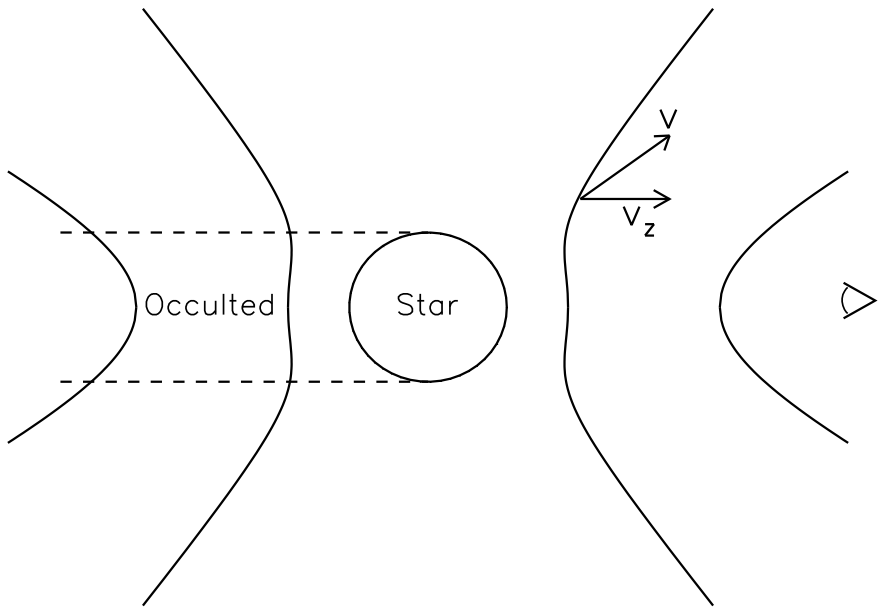}
\caption{Shown are ``iso-velocity surfaces'' in a wind, on which
the component of velocity toward the observer, $V_z$, is constant.
The net scattering along the each surface both into and out of the line 
of sight gives the observed flux at a given wavelength in the line. }
\end{figure}

In the iso-velocity approach, first developed by Chandrasekhar (1934) and
extended by Sobolev (1947), we consider one wavelength band of the profile
at a time. The flux at a Doppler displacement, $\Delta
\lambda$ = $\lambda$ $V_z/c$, from line center is formed along one of the
constant velocity surfaces shown in figure 2. In the shortward shifted side
of the line, there is both a reduction of the light because of scattering by
ions in the area in front of the star's disk, and an increase by the
scattering toward the observer from the entire constant velocity region. The
iso-velocity approach shows that we can resolve sectors of the envelope
in wavelength space; that is sufficient for applying the Hanle effect.
Now, let us include in this picture a magnetic field in the wind. For this
discussion, consider a spherically symmetric wind with a superposed
stellar dipole field. The spherically distribution for the Rayleigh
scatterers would lead to zero net polarization.  However, the dipole field
introduces an {\it asymmetry} needed for the Hanle effect. At every point in
our model envelope the Hanle source function can be calculated and then
integrated to find the monochromatic polarized line flux.

The fiducial magnetic field for a given line is the Hanle Field, $B_H (
\approx 5$ Gauss $ \times \frac{A_{u,l}}{10^8}$), where A$_{ul}$ is the
decay rate for the $u$ to $l$ transition.  Figure 1 in Nordsieck's paper
(these proceedings) gives polarizability results for the UV features of
interest here. Many are closely separated doublets, with each component
having a different $E_1$ value. Typically, one line of the doublet is sensitive
to the magnetic field and the other, an isotropic scattering line, is not.
An inter-comparison of the two components is useful for isolating the magnetic
effects.

Figure 3 shows a simulation of an observation of $\zeta$ Ori A, a supergiant
for which there is indirect evidence for the presence of a magnetic field in
that the star emits X-rays corresponding to an anomalously hot gas. (Cassinelli
and Swank, 1983). In Figure 3 are shown clearly observable changes in the line
polarizations of the resonance lines shown as the surface field is increased
from 1 to 100 Gauss.

\vspace{-.5cm}
\begin{figure}[ht]
\plotone{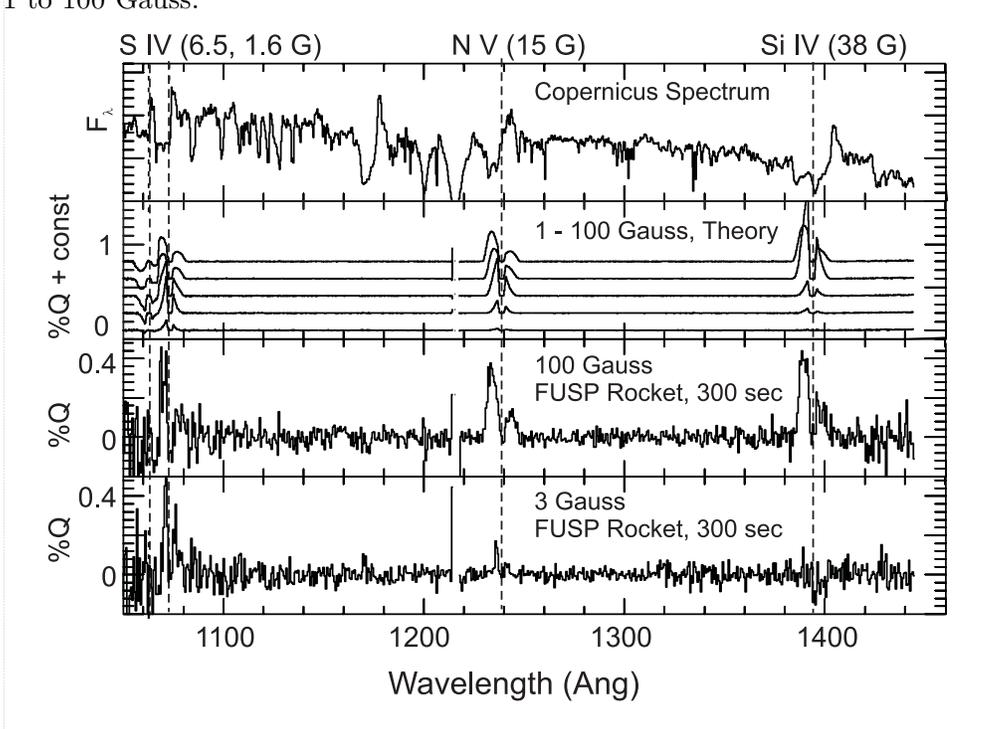}
\caption{Numerical simulation of the Hanle Effect for the case of a
dipole field in a spherically symmetric wind. The top panel shows the 
{\it Copernicus} UV spectrum of $\zeta$ Ori (O9.5 Ia). Four lines with 
the given Hanle fields are noted at the top. The next panel
shows the theoretical polarized profiles at surface field strengths of
1, 3, 10, 30, and 100 Gauss. The lower two panels show a simulation of
a 300 second FUSP exposure of the star with surface field strengths of 
3 and 100 Gauss.}
\end{figure}

\section{Contrasts with the Solar Case}

We have heard several other papers at this conference about the Hanle effect
as it is applied to the Sun. Let us consider some of the problems raised.
\\[.1cm]
A) Landi Degl'Innocenti stressed the the Hanle effect is a non LTE effect.
For the solar application this implies a complicated multilevel atom
radiation transfer problem. For our case, we can use the much simpler
nebular approximation that all atoms/ions are in their ground states, as we
are dealing with resonance scattering starting from the $n=1$ level.\\[.1cm]
B) Fineschi stated in his presentation that the presence of velocity fields
limits the applicability of the Hanle effect.  This statement would seem to
be a major concern for hot stars, as we are always dealing with velocity
fields-- the winds. However, Fineschi has explained that for the solar case
the incident light that is scattered is Chromospheric line emission. Thus
the incident flux varies strongly with wavelength across the profile.
Transverse motions of the zone being diagnosed produce Doppler shifts
relative to opposing limbs of the Sun. This leads to an asymmetry in the the
scattered radiation, and to what is called a ``false'' Hanle Effect.
However, for the hot stars this is not a problem because the light that is
scattered in the envelope is {\it flat} continuum emission from the star,
ie. $\frac{d\Flam}{d \lambda} \approx 0$, over the wavelengths where each
Hanle diagnostic line forms.\\[.1cm] 
C) Landi Degl'Innocenti also said in his review that
Hanle is a useful diagnostic over only a small range in magnetic field, and
showed a plot of polarization versus B to support that statement. However,
for the UV one could show several such plots, one for each of the Hanle
sensitive lines, and conclude that Hanle will be useful to derive magnetic
fields in the range 1 to 300 Gauss. Coincidently, this range corresponds
well with the values in surface magnetic fields that can have significant
dynamical effects on the wind, (Cassinelli and Miller, 1999). For the
Be-stars, the disks could be magnetically spun-up by magnetic fields of
about 30 gauss. It is a also range that is difficult to measure with the
Zeeman effect owing to the immense line broadening.

\section{Conclusions}

The Hanle Effect offers a promising way to derive the magnetic fields on
early-type stars. The lines needed for the field diagnostics are among the
most prominent lines in hot star spectra. With P-Cygni lines, we can derive
spatial information about the magnetic field, since each segment of the
line profile is formed in a different isovelocity sector of the wind.
Three problems affecting the application of the Hanle effect to the Sun
are expected to be absent in applications to hot stars.

\newpage

\section*{Discussion}

\noindent BUENO: I have a question with respect to the UV lines you are
proposing for the Hanle effect diagnostics of magnetic fields in the winds
of hot stars. Can you really model their Stokes profiles by assuming that
they are optically thin in the stellar wind?\\[.1cm] 

\noindent CASSINELLI: Although resonance line cross-sections are large, the
optical depths in lines can be moderate because of the expansion motions of
the wind. The relevant optical depths are those through narrow interaction
regions or Sobolev zones, where $\tau$ depends inversely on the velocity
gradient. Also, because of the ionization fractions involved, there are
lines in the UV with a range of strengths. The lines of $\zeta$ Ori (Figure
3) have P-Cygni profiles that are not saturated and indicate optical depths
of order unity or less across the lines. It is lines of intermediate
strength, i.e. moderate optical depths, that have proven to be of greatest
value in nearly all line profile diagnostic studies of hot stars, and the
same will hold true for the Hanle effect studies.


\begin{references}
\reference Cassinelli, J. P. \& Miller, N. A. 1999, in {\it Variable and Non
Spherical Winds in Luminous Hot Stars} Eds. B. Wolf et al. 
(Berlin,Springer)p 169.
\reference Cassinelli, J. P. \& Swank, J. H. 1983, \apj, 271, 681
\reference Chandrasekhar, S. 1934, \mnras, 94, 522
\reference Ignace, R., Nordsieck, K. H., $\&$ Cassinelli, J. P. 1997, 
      \apj, 486, 550 (Paper I)
\reference Ignace, R.,Cassinelli, J. P., $\&$  Nordsieck, K. H.  1999, 
      \apj, 520, 335 (Paper II)
\reference Ignace, R. 2001, \apj\ 547, 393 (Paper III)
\reference Lamers, H. J. G. L. M. \&  Cassinelli, J. P. 1999
{\it Introduction to Stellar Winds} Cambridge University Press.
\reference Sobolev, V. 1960 {\it Moving Envelopes of Stars} 
      (Cambridge, MA: Harvard Univ. Press) [Russian Edition 1947] 
\end{references}
\end{document}